%
\def\lsim{\, \lower2truept\hbox{${< \atop\hbox{\raise4truept\hbox{$\sim$}}}$}\,}
\def\gsim{\, \lower2truept\hbox{${> \atop\hbox{\raise4truept\hbox{$\sim$}}}$}\,}
 
\def\ms{\, {\rm M_\odot}}
\def\fir-rad{L_{\rm FIR}/P_{\rm rad}}
\def\lfir{L_{\rm FIR}}
\def\prad{P_{\rm rad}}
\def\ub{U_{B}}
\def\psyn{P_{\rm syn}}
\def\p24{P_{\rm 2.4 GHz}}
\def\ptherm{P_{\rm therm}}
\def\urad{U_{\rm rad}}
\def\q{{\nu_{SN} q_{SN} \over 2D_0}}

\def\hii{H\,{\sc ii}}
\def\lhii{L_{\rm FIR}^{\rm HII}}
\def\ldiff{L_{\rm FIR}^{\rm diff}}
\def\lb{L_{\rm B}}
\def\mlow{m_{\rm l}}
\def\nlyc{N_{\rm Lyc}}

\def\q24{q_{\rm 2.4GHz}}
\def\tauloss{\tau_{\rm loss}}

\def\tausb{\tau_{\rm SB}}

\def\taugrow{\tau_{\rm grow}}

\def\apj{ApJ }

\def\aa{A\&A }
\def\aas{A\&AS }

\def\mnras{MNRAS }
\def\araa{A\&AR }
\def\aj{AJ }
\def\figtext{\figure}
\def\figonea{\begfig 8.cm}
\def\figoneb{\begfig 8.cm}
\def\figtwo{\begfig 8.cm}
\def\figthree{\begfig 8.cm}
\def\figfour{\begfig 8.cm}
\def\figfive{\begfig 8.cm}
\def\figsix{\begfig 8.cm}
\def\figseven{\begfig 8.cm}
\def\figeight{\begfig 8.cm}
\def\fignine{\begfig 8.cm}
\def\figpos{ }
%
\input cp-aa.tex
%
%
  \MAINTITLE={The FIR/radio correlation in starburst galaxies -- 
              constraints on starburst models}
%
%
%
%
  \AUTHOR={U.~Lisenfeld@*, H.J.~V\"olk, C.~Xu} 
                             \PRESADD{Osservatorio Astrofisico di
			     Arcetri, Largo E. Fermi 5, I-50125
			     Florence, Italy}
  \INSTITUTE={Max-Planck-Institut f\"ur Kernphysik, Postfach 10 35 80,
  D-69029 Heidelberg, Germany}
%
  \ABSTRACT={
This paper presents an analysis of 
the correlation between the far-infrared (FIR)
and the radio emission of  starburst galaxies.   
Data for interacting galaxies, many of which
are undergoing a starburst, and for normal galaxies have
been analysed and compared in order
to test for any influence of the star-formation activity on
the ratio between the 
FIR and the radio emission at 2.4 GHz, $\q24$. 
No statistically significant indication for such an influence
has been found:
There is neither a significant 
difference between $\q24$ of the two
samples nor a dependence of this
ratio on the starburst strength.
This observational fact is unexpected
because of the different physical conditions and
the short time-scale of the star-formation activity in a starburst.
In order to interprete the observations
the FIR and the radio emission during a starburst have been modeled.
The following conclusions could be drawn from the observed 
constancy of $\q24$: 
A strong and fast ($\approx 10^7$ yr)
increase of the magnetic field at the beginning of the
starburst is required in order to maintain a constant
$\q24$.
Otherwise the strong Inverse Compton losses
that are due to the intense radiation field in a starburst
would lower the  synchrotron emission drastically 
resulting in a value of $\q24$ significantly higher than
the observed one.
Furthermore,
the time-scale of the 
variation of the star-formation rate  has to be  
longer than some $ 10^7$ yr. For lower values the
different time-scales of the FIR and the radio emission
produce  large fluctuations of $\q24$.
}
 \KEYWORDS={Galaxies: starburst -- galaxies: magnetic fields --
 infrared: galaxies -- radio continuum: galaxies }       
  \THESAURUS={03(11.13.2; 11.19.3; 13.09.1; 13.18.1)}      
\OFFPRINTS={H.J.~V\"olk}      
  \DATE={Received 30 January 1996/ Accepted 22 April 1996}           
%
\maketitle
%
%
\titlea{Introduction}
Starburst galaxies are found to follow the 
tight correlation between the far-infrared (FIR) and the radio
emission  
(Wunderlich et al. 1987; van den Driel et al. 1991; Franceschini et al. 1988a;
Condon et al. 1991).
Here, the correlation is even more
unexpected  than in normal galaxies
(see Lisenfeld et al. 1995 for a discussion
of normal galaxies) because of the following reasons. 
In starburst galaxies 
the different time-scales 
involved in the
production of the radio and the FIR emissions become important:
Whereas the FIR emission and the thermal
radio emission are instantaneously related to the
stellar radiation, the synchrotron
emission shows a time delay because it takes 
some $10^6$ yr until the first supernovae (SNe)  
occur, and some $10^7$ yr till the maximum of SN activity is 
reached.
These time-scales are comparable to the duration of starbursts 
of about $10^7-10^8 $
yr (Rieke et al. 1980; van den Broek et al. 1991; Bernl\"ohr 1993)
and should therefore result in a time delay between the maximum
emission of the FIR and the radio emission.
Furthermore, 
during a starburst the intensity of the radiation field
changes with time in a significant way. Little is known about 
variations of the magnetic field energy density. 
The ratio of these two parameters that determine the major
radiative losses of  the cosmic ray (CR) electrons (inverse Compton and 
synchrotron losses) is crucial for  the total
synchrotron emission of a galaxy.
\par
The aim of this paper is to understand why,
in spite of the above considerations, the FIR/radio ratio
is observed to be very constant in starburst galaxies,
and, more importantly,
to use this observational fact 
to find constraints for  starburst models.
The paper consists of two parts: In the first part (Sects. 2 and 3), 
we present data for two large samples of galaxies, a starburst sample
and a control sample, in order to investigate 
whether the FIR/radio ratio shows any dependence on the
star-formation activity.
In the second part (Sects. 4-6) we
present a model for the FIR and the radio emission in a starburst
in order to interprete the observational results. Section 7 summarizes
our conclusions.
\par
\titlea{The sample}
Both the sample of starburst galaxies and the control sample are selected from
the list of UGC galaxies which are surveyed with the Arecibo 300m telescope
at 2.4 GHz (Dressel \& Condon 1978) and were covered by IRAS
in FIR . These galaxies form a
complete sample (down to $\rm m_B =14.5$)
of a local galaxy population, which have been studied
by Franceschini et al. (1988a,b) for the radio and FIR
luminosity functions.

It has been well established that non-Seyfert Markarian galaxies
(Huchra 1977; Balzano 1983; Xu \& De Zotti 1989;
Mazzarella et al. 1991)
and interacting galaxies (Larson \& Tinsley 1978;
Kennicutt et al. 1987; Telesco et al. 1988; Xu \& Sulentic 1991)
have on average significantly higher current
star-formation rates than standard galaxies like the Milky Way.
Therefore we select these galaxies for our starburst galaxy sample.
It includes those galaxies classified
in the UGC catalog (Nilson 1973) as peculiars, double, trible, or
multiple systems, and galaxies identified by Markarian and collaborators as
having ultraviolet excesses (see Mazzarella \& Balzano 1986 for a review).
Galaxies with active galactic nuclei (e.g. Seyfert galaxies) are 
not considered. The region of low
galactic latitude ($\rm |b|< 30^\circ$) is avoided in order
to minimize the influence of Galactic extinction.
Virgo galaxies listed in the VCC catalog (Binggeli et al. 1985)
are excluded, because  in clusters galaxies may have additional
radio emission mechanisms other than the star-formation alone 
(Gavazzi et al. 1991; V\"olk \&  Xu 1994).
There are 133 objects in the starburst galaxy sample.

The control sample includes spiral galaxies which are at high Galactic
latitude ($\rm |b|> 40^\circ$), 
and from the morphological types Sa through Sd in the UGC catalog.
The sample excludes Virgo cluster, Markarian, as well as Seyfert
galaxies.
Galaxies catagorized simply as 'S' are not included, because the
probability that they are misclassified 
early types is relatively high.
There are 397 spirals in the control sample.

For the majority of objects in both samples, fluxes at 60$\mu \rm m$
and 100$\mu\rm m $  are taken from
the Second Version IRAS Point Source Catalog. The exceptions are those extended
(size $> 5'$) and faint ($f_{60\mu {\rm m}} < 0.6$ Jy) sources listed in
the catalog of IRAS Faint Source Survey
(FSS) fluxes of CfA galaxies (Thuan \&  Sauvage, 1992),
for which the data are taken from Thuan \&  Sauvage (1992).
For 133 objects in the starburst sample, 113 are detected at
both $60 \mu \rm m $ and $100 \mu \rm m $, 
corresponding to a detection rate of 85\%.
For the control sample, 329 are detected out of 397 objects,
i.e. a detection rate of 83$\%$. The integrated flux
in the wavelength range 42.5 --- 122.5 $\mu$m, $S_{FIR}$  , 
or its upper limit,
is calculated from the IRAS fluxes (limits) according to Helou et al. (1988).
Radio fluxes at 2.4 GHz
are taken from the Arecibo survey data (Dressel \& Condon
1978).
Fourty-six objects in the starburst sample
and 92 in the control sample are
detected in radio, corresponding to detection rates of 35\% and 23\%,
respectively. For an overwhelming majority of sources in our samples,
the total ``face-on" blue magnitude $\rm B^0_T$, corrected for Galactic 
extinction, inclination effect and for redshift, can be found in the Third
Reference Catalog (RC3) (de Vaucouleurs et al. 1991). For a few exceptionals
whose $\rm B_T^0$ are not available, the blue magnitude is simply taken from
the photographic magnitude given in UGC and corrected for the Galactic
extinction calculated using the formula given in RC3.
Taking $\rm H_0= 75 km s^{-1} Mpc^{-1}$, 
the distances are calculated from the recession velocities
after correcting for the Virgocentric flow (Aaronson et al. 1982).
\titlea{Statistical results}
\figonea
\figpos
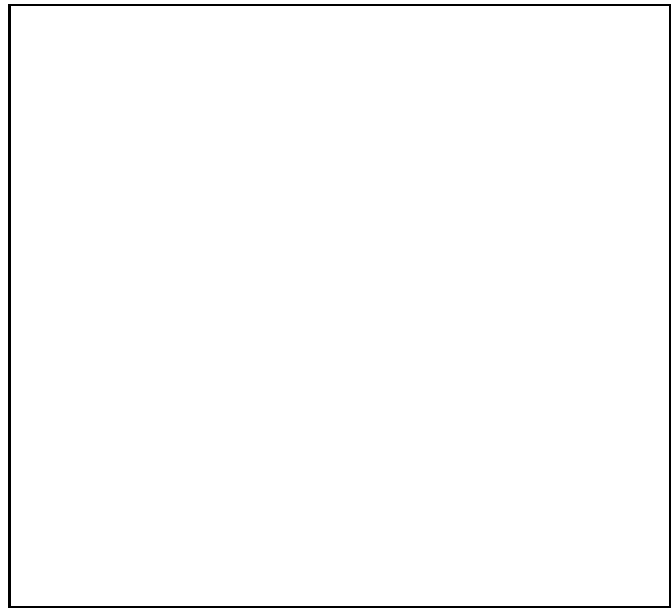
\figtext{1a}{The FIR and the radio luminostities for the starburst
sample. Upper limits are indicated by arrows.}
\endfig

In the statistical analysis carried out in this section,
we use the so called `survival technique' (Schmitt 1985;
Feigelson \& Nelson 1985) to exploit the information carried by the
upper limits.

We plot in Figs. 1a and  1b the
logarithms of the radio intensities 
versus the logarithms of the FIR luminosities
for the starburst sample and for the control sample, respectively.
The crosses are detected data, and the arrows denote the upper limits.
Both samples show tight correlations between the FIR and the radio
luminosities: the linear correlation coefficients are 0.90 and 0.88
for the starburst sample and the control sample, respectively.
The slope of the log-log plot of the starburst sample is very close to unity:
0.98$\pm 0.01$. On the other hand, the radio intensities increase
slightly faster than the FIR luminosities for
the normal spiral galaxies in the control sample:
the slope is equal to 1.09$\pm 0.01$ in Fig. 1b. In both plots,
no galaxy is detected to seriously violate the correlation.
\figoneb
\figpos
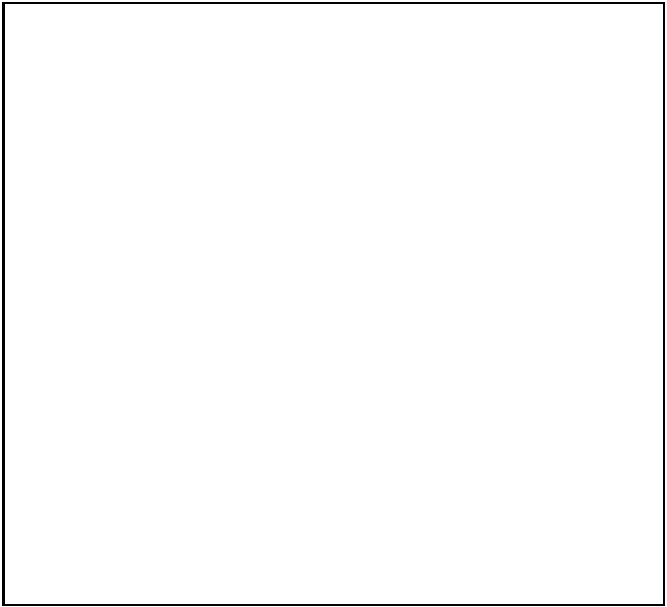
\figtext{1b}{The FIR and the radio luminostities for the control
sample. Upper limits are indicated by arrows.}
\endfig
Similar to the $q$ parameter defined by Helou et al. (1985), we introduce
a parameter $\q24$:
$$\rm \q24 = \log [(S_{FIR}/3.75\; 
10^{12})/S_\nu (2.4GHz)] \eqno\autnum $$
For the starburst sample, the
mean of the $\q24$ is $2.40\pm 0.03$, while for the control sample
it  is $2.52\pm 0.02$. 
The statistical dispersion of the logarithm of the FIR/radio ratio, calculated
using the survival technique, is $\sigma=$ 0.22 for the starburst
sample, and $\sigma=$ 0.21 for the control sample. 
Therefore, although the $\q24$ value of the starburst sample is
slightly lower than the control sample, the difference is within 
the dispersion.

An important question in this study is: does the FIR/radio ratio depend
on the starburst strength? 
In principle, the starburst strength can be defined either via the
FIR--to--blue luminosity ratio or via the radio--to--blue luminosity
ratio. Both reflect, approximately, the ratio between the star-formation
rates averaged over the last $\sim 10^8$ yr, $s_8$, and the
last $\sim 3 \times 10^9 $ yr, $s_9$, respectively (Xu et al. 1994). 
However, there is a
trivial dependence (positive) of the FIR--to--radio ratio on the
FIR--to--blue ratio, because the FIR luminosity is involved in both
ratios. Similarly, a trivial dependence (negative) exists
between the FIR--to--radio ratio and the radio--to--blue ratio. In order
to minimize this artificial effect, we introduce a starburst
strength indicator SB invoking both ratios:
$$\eqalign{\rm SB = & \rm 0.5\times \lbrack \left( \log{ \left(
                   {\lfir\over L_B} \right)} \
  \; -\; \langle \log{\left( {\lfir \over L_B} \right)} \rangle \right)\; + \cr
 & \rm \left( \log{ \left( {\p24\over L_B} \right)} \; -\;  \langle
  \log{\left( {\p24 \over L_B} \right) } \rangle \right) \rbrack }
       \eqno\autnum $$
where $\rm \langle \log \left({\lfir\over L_B} \right) \rangle=
-0.47\pm0.02$
and  $\rm \langle\log \left( {\p24 \over L_B}\right) \rangle=
-15.70\pm 0.05$ 
are the means of the
logarithms of the FIR--to--blue and of the radio--to--blue luminosity ratios
of the {\it control} sample. 
A mean SB of 0.46$\pm 0.05$ is found for the starburst sample, indicating
that on average the $s_8$--to--$s_9$ ratio
is a factor of 2.9$\pm 0.4$ higher than that of the normal spirals in
the control sample,
consistent with previous studies (Kennicutt et al. 1987; Xu \& Sulentic 1991).

%
\figtwo  
\figpos
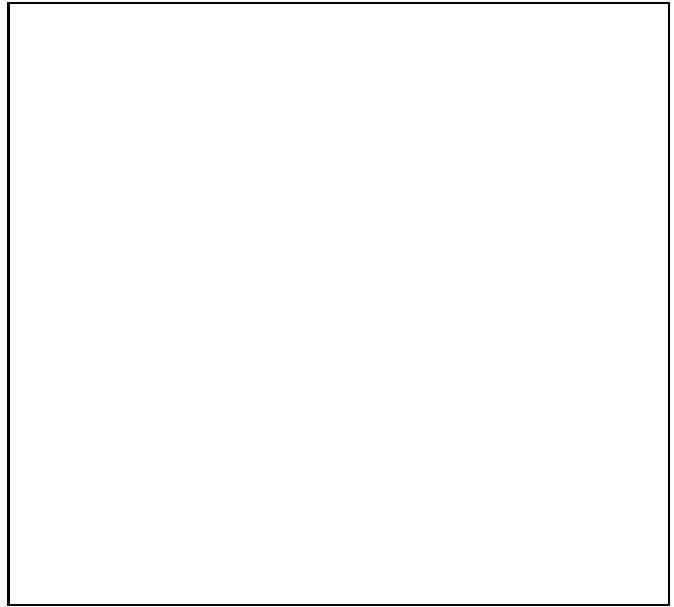
\figtext{2}{The logarithmic ratio of the FIR--to--radio luminostity, $\q24$,
and the indicator of the starburst strength, SB, defined in Eq. 2,
for the starburst sample}
\endfig
\figthree
\figpos
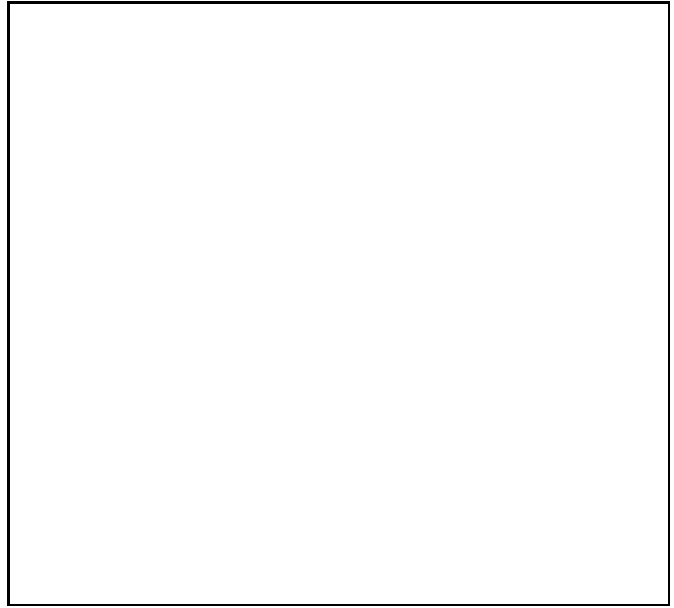
\figtext{3}{$\q24$ and the logarithm of the FIR colour, 
$\log f_{60\mu \rm m}/f_{100\mu \rm m}$, for the starburst sample}
\endfig
Fig. 2 shows    $q_{2.4GHz}$ vs SB  for 113 objects in
the starburst sample. The other twenty sources 
undetected either in the FIR or in the radio are dropped
because no constraint can be imposed on their $\q24$ values.
No correlation is found in this diagram. The two galaxies
with the highest SB values show very normal $\q24$.
The variation of the
FIR--to--blue and the radio--to--blue ratios, so the
SB, may be due to the variation
of the extinction, in addition to the depence on the starburst strength.
In order to verify the results obtained in Fig. 2, we plot in Fig. 3
the dependence of $\q24$ on another -- extinction-free --
starburst indicator, namely
the $\log f_{60\mu m}/f_{100\mu m}$, for the same galaxies as in
Fig. 2. As pointed out by Xu \&  De Zotti (1989)
and by Bothun et al. (1989), $\log f_{60\mu m}/f_{100\mu m}$ is tightly
associated to the high mass star-formation rate of galaxies, and 
the galaxies
with the strongest starbursts usually show the highest
$\log f_{60\mu m}/f_{100\mu m}$ values. Again, we do not detect any
dependence of $\q24$ on this starburst indicator.
Therefore, we conclude that
the strength of the starburst does not introduce any siginificant
influence on the FIR--to--radio ratio.
\titlea{The starburst model}
\titleb{Star formation during a starburst}
In the model we describe the starburst  galaxy 
as the sum of two components: a starburst population with
a star formation rate (SFR) $\Psi_{\rm SB}(t)$ 
changing on short time-scales and a background population
whose SFR $\Psi_{bg}$ is constant in time. The total SFR is the
sum of both components, $\Psi(t)= \Psi_{\rm SB}(t)+\Psi_{\rm bg}$.
The SFR of the starburst population is parametrized in the following way:
$$ \Psi_{\rm SB}(t)=\cases{\Psi_0 \, \exp^{t/\taugrow} &{\rm for} $t<0$  \cr
                  \Psi_0\,\exp^{-t/\tausb} & {\rm for} $t\ge0$ \cr}
                        \eqno{\autnum}$$
where $\taugrow$ and $\tausb$ are the time-scales of the increase and the
decrease of the SFR.
The ratio between the maximum SFR of the starburst population and
the background population
$a=\Psi_0/\Psi_{\rm bg}$ is a measure of the 
starburst  strength. An upper limit for $a$ can be estimated from
the total amount of molecular gas available. The SFR of the background 
population can be estimated by assuming that the total amount of
gas present is used up in about 
$10^{10}$ yr. Then we can estimate an upper limit of $a$ by:
$$ a=\Psi_0/\Psi_{\rm bg}\le {10^{10}\cdot x 
\over \tausb }\eqno\autnum$$
where $x$ is the fraction of molecular gas available for the 
starburst. In a 
nuclear starburst, about $20-30\%$ of the
total amount of gas is in the nucleus
and therefore available for
the starburst (Bernl\"ohr 1992). The efficiency of the conversion
of gas to \nobreak{stars} can reach in starburst galaxies  $50\%$ 
(e.g. Rieke et al. 1980).
With these numbers we derive $x=0.1-0.15$. Thus, e.g., 
for $\tausb=10^7$ yr, 
an upper limit for $a$ is about 100.
\par
The Initial Mass Function (IMF), $\Phi(m)$, 
which determines the mass distribution
of the stars produced, is described by a power-law
$$\Phi(m)\diff m=m^{-\gamma}\diff m \eqno\autnum$$
with an upper mass limit $m_u=100\ms$ 
and a lower mass limit $\mlow= 1 \ms$ 
(stars less massive than that are neither of importance for 
the FIR nor the radio emission).
We adopt $\gamma=2.7$,
in agreement with Scalo (1986) and
the more recent results of Basu and Rana (1992). In addition, 
we  considered 
the effect of changes in the slope of the IMF and $\mlow$.
\par
\titleb{FIR emission}
The FIR luminosity is calculated as
$$\lfir(t)=\int_{m_l}^{m_u} \diff m \Phi(m) L(m) P(m) \zeta(m)
\int_{t-\tau(m)}^t \diff t 
\Psi(t)\eqno\autnum $$
where $\tau(m)$ is the stellar main-sequenze life-time,
$L(m)$ the stellar luminosity
(taken from 
Cox et al. 1988 and Cox et al. 1986),
$P(m)$ the fraction of light absorbed
by the dust and $\zeta(m)$ the fraction of the absorbed light that
is reemitted in the IRAS 40--120 $\mu$m band. 
Here, we parametrize the last two parameters in the
same way as explained in detail in 
Lisenfeld et al. (1995):
For massive, ionizing stars ($m>20 \ms$) we assume $P(m)=0.6,
\zeta(m)=0.5$. The radiation transfer of the non-ionizing 
emission of the intermediate massive stars ($5 \ms<m<20 \ms$) and
the old stars ($1\ms<m<5 \ms$) is approximated by an infinite
parallel-slab geometry in which
dust is homogenously mixed with the stars. We assume $A_V=1.5$ for the
starburst population (Bernl\"ohr 1993) and $A_V=0.2$ for the
background population (Lisenfeld et al. 1995; Xu \& Buat 1995).
We make the further simplification that the radiation of 
the intermediate massive stars 
is absorbed according to the dust opacity in the UV (2000 \AA)
and the radiation of the old stars according
to the dust opacity in the blue (4400 \AA). 

Finally we split the FIR luminosity
into the emission coming from
\hii \  regions, $\lhii$, and the diffuse emission $\ldiff$, by
assuming that stars more massive than 20 $\ms$ are responsible for
the dust heating in \hii \  regions, and less massive stars for 
the diffuse dust heating.

\titleb{Radio emission}

The radio emission consists of thermal bremsstrahlung
and of synchrotron 
emission.
The thermal radio emission is proportional to
the total number of Lyman continuum photons $\nlyc$ emitted by the stars:
$$\ptherm(\nu)=
 1.08\cdot 10^{-33}
\biggl({\nlyc \over \rm phot \, s^{-1}}\biggr) \biggl({ \nu\over {\rm GHz}} 
\biggr)^{-0,1}
[{\rm W/Hz}]\eqno\autnum$$
(e.g. Condon 1992)
where we have assumed that the fraction of Lyman continuum
photons directly absorbed by the dust is $f=0.2$ and
the electron temperature in the \hii \ region
$T_{\rm e} =7000$ K.
The total number of Lyman continuum photons is calculated as:
$$ \nlyc= \int_{\mlow}^{m_{\rm u}}\diff m \phi(m) \tilde n_{\rm lyc}(m) 
\int_{t-\tau(m)}
^t \diff t \Psi(t) \eqno\autnum$$
where $\tilde n_{\rm Lyc}(m)$ is the production rate of Lyman continuum
photons of a star of mass $m$, taken from G\"usten \& Mezger (1983).

In order to calculate the synchrotron emission in a starburst a
time-dependent equation has to be solved.
The details are described in the Appendix. Here we will only 
outline the most important features.
Several
processes are taken into account:
We assume that the CR electrons are accelerated in shocks of SN 
remnants.
Then they propagate into the galactic disk and halo losing their energy
radiatively through inverse Compton and synchrotron losses.
Since the radiation field in a starburst is very strong,
the inverse Compton losses of the electrons are very high. This has the effect
of shortening their radiative life-time, $\tauloss$, 
and therefore of decreasing the temporal shift between the
FIR and the synchrotron radiation.
(In a radiation field which is, for example, a factor of 10 higher
than in the Galaxy, i.e. about 10 eV cm$^{-3}$, 
the life-time is 
$\tauloss\approx 10^6$ yr.)
On the other hand, a high radiation field lowers
the synchrotron emission itself, because the number density 
of electrons at
a given energy is lowered (see Appendix).
Therefore, during a
starburst -- if the magnetic field stays constant -- the
synchrotron emission  is a sensitve function of the radiation field and
therefore of the starburst strength. In  particular it is very low
during
the phase of most active star-formation.
\titleb{Starburst strength parameter SB}
In order to compare our model with the data displayed in Fig. 3,
we compute in the model the starburst parameter \nobreak{SB}.
The blue  luminosity is  obtained by assuming that the stellar
emission is  a blackbody spectrum. 
We obtain for
the background, steady state population of the galaxy
within our model:
$\langle \log{\left( {\lfir\over \lb} \right)} \rangle =-0.34 $ and 
$\langle
  \log{\left( {\p24\over \lb} \right) } \rangle = -15.48$.
The model predictions 
are in rather good agreement with the mean values  of the control
sample. The small difference between model and data of   
$\langle \log{\left( {\lfir\over \lb} \right)} \rangle$,
might be due to a slight overestimate of
the average galactic extinction or due to the crude model assumptions
(e.g. blackbody emission of stars).
In the model $\p24$ for the background population 
is calculated from the model result for $\lfir$ and 
the mean value of $\q24$ of the control sample. Therefore 
the model prediction
for
$\langle
  \log{\left( {\p24\over \lb} \right) } \rangle $
is simply a combination of 	
$\langle \log{\left( {\lfir\over \lb} \right)} \rangle$ (model),
and $\q24$ (control sample). 
\par
The value of SB depends very sensitively on the extinction.
The extinction in the starburst region  is most likely higher than
for the background population. We assumed  
$A_V=1.5$, 
which is 
the typical internal extinction of the \hii \ regions in nearby galaxies
(Kennicutt 1983), and at the same time consistent with the average
$A_V$ found for the Markarian galaxies by Xu \& De Zotti (1989).

In reality it seems plausible that some galaxies have an even higher  
extinction, explaining the large range in
SB found for the starburst sample.
The extinction
mainly has an influence on SB,
through $L_B$,
and hardly on the FIR emission (see 
Fig. 9) because
even for the value of $A_V$ assumed for the background population
($A_V=0.2$)
the galaxies are optically thick for the
nonionizing UV light which is the main heating
source for the dust (Xu 1990). 
\par
\titlea{Variation of $\q24$ during a starburst}
In order to show  the general features of the temporal evolution
of $\q24$ during a starburst, we display in Fig. 4 
the variation of $\q24$ during a starburst with a timescale
$\tausb=10^7$ yr and $\taugrow=0$ for different starburst strengths,
and in the Figs. 5 and 6 separately the temporal
evolution of the FIR and the radio emission.
\figfour
\figpos
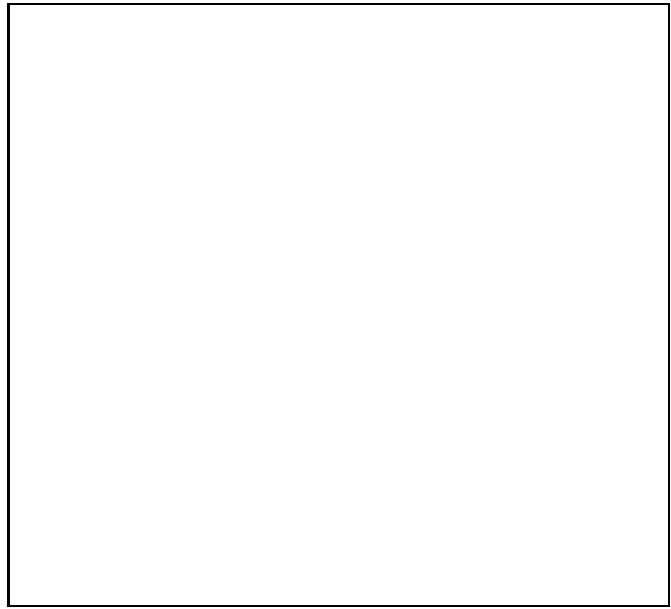
\figtext{4}{The ratio between the FIR and the radio emission at 2.4 GHz,
$\q24(t)$,
is shown as a function of time for $\tausb=10^7$ yr, 
$\taugrow=0$, $B=$const, $a=\Psi_0/\Psi_{\rm bg}
=100$ (full line), $a=50$ (dashed line), and $a=25$ (dotted line).
The dash-dotted line
shows $\q24$ for a steady state galaxy.}
\endfig
\figfive
\figpos
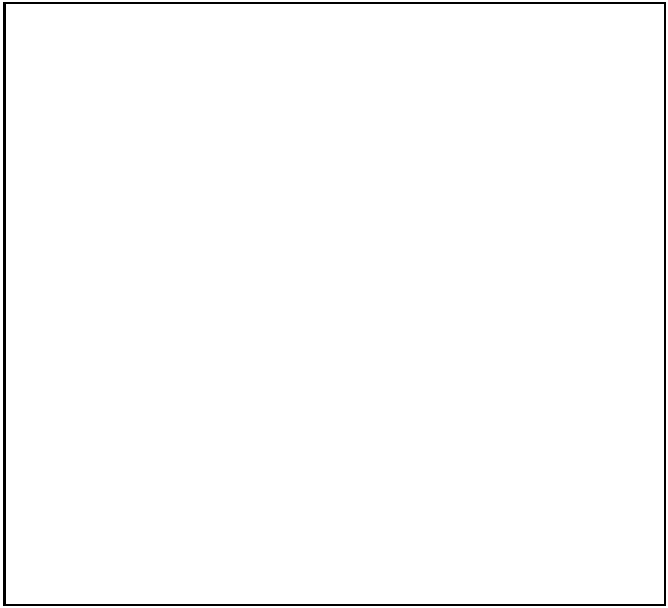
\figtext{5}{For a starburst with  $a=100$ and otherwise the
same parameters as in Fig. 4 the
FIR emission (full line) is shown as a function of time. The FIR emission from
\hii \ regions (dashed line) and the FIR emission from diffuse dust (dotted
line) are shown separately. The dash-dotted line describes the FIR emission
from the background galaxy.}
\endfig
The temporal variation  of $\q24$ can be split into 3 phases:
In the early phase (phase I; here: until $\approx 50$ Myr) 
$\q24 (t)$ is higher than
the steady state value at this frequency.
In this phase the emissions from \hii \ regions
dominate the total FIR and radio emission (see Figs. 5 and 6).
With respect to the FIR emission, this is due to
the  high relative abundance of massive stars. For the synchrotron 
emission other reasons are more important: First of all, it
takes serveral $10^6$ yr until  the first
SN explosions occur. More important, however, is the
fact that during the first serveral $10^7$ yr the synchrotron emission
is strongly suppressed (see Fig. 6) due to the  
strong inverse Compton losses caused by a high radiation field.
It is worth to note that only because  of the presence of the thermal
radio emission and because the two ratios 
($\log(\lhii/\ptherm)$ and $\log(\ldiff/\psyn)$ are
relatively similiar 
no major difference of $\q24$ with respect to a steady state galaxy occurs
in this phase.
\par
\figsix
\figpos
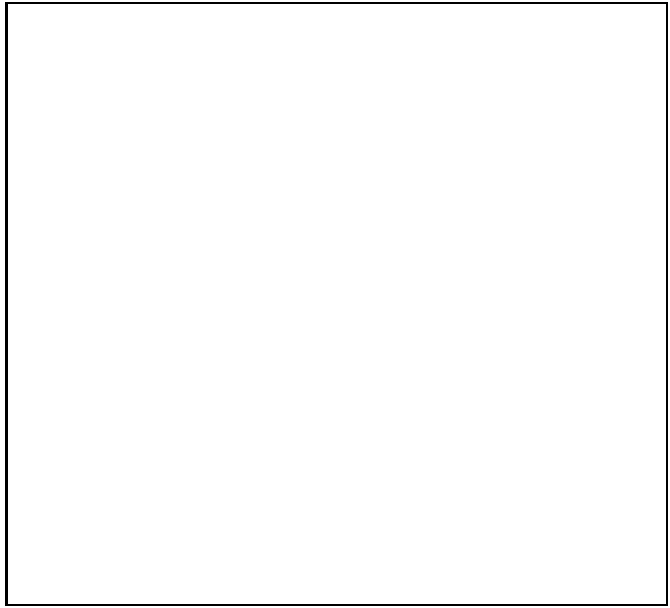
\figtext{6}{For $a=100$ and otherwise the same parameters as in Fig. 4 
the radio emission during
the starburst is shown (full line), devided into thermal bremsstrahlung
(dashed line) and synchrotron emission (dotted line). The dash-dotted line
describes the radio emission from the background galaxy.}
\endfig
In the next phase (phase II; here: between about $50-100$ Myr),
$\q24$ can be even lower than in a steady state.  This is 
due to the time delay between the SN explosions and the FIR
emission: it  leads to a maximum of the 
radio emission at a time when the FIR emission is already in its
declining phase. 
\par
In the last phase (phase III; here: after about 100 Myr)
the opposite temporal effect occurs:
The synchrotron radiation has already decreased  to the
steady state value -- or even below because the inverse
Compton losses are still higher than before the starburst -- but the
FIR emission is still high due to dust heating by low-mass
($m=1-3\ms$)
stars. Therefore during this phase $\q24$ is  higher than the
steady state value and  depends sensitively on
$a=\Psi_0/\Psi_{\rm bg}$. In this phase the galaxy is already
completely in the post-starburst phase. 
\par
\figseven
\figpos
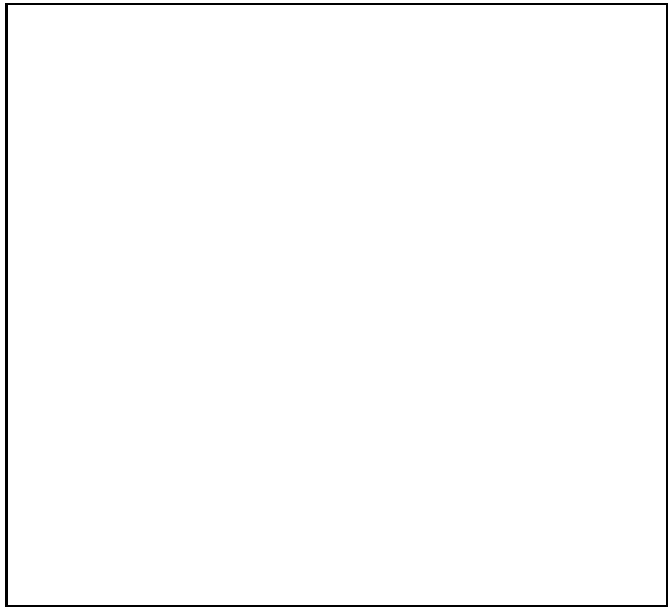
\figtext{7}{The model result for the same parameters as in Fig. 4
(with $a=100$) are compared to the data. 
The full line shows the model results for SB and $\q24$
for the time interval
between $t=0$ and $t=10^9$ yr with the arrow indicating
the direction of increasing time.}
\endfig
The deviations of $\q24$ from the steady state value 
in the various phases are therefore either due
to the different time-scales of the 
FIR and the radio emission or due to the high inverse Compton losses
experienced by the electrons.

In Fig. 7 the model results are compared to the data. It can be seen
that in phase I  the model predictions for $\q24$ lie
substantially above  the data.
In the following we will try to find parameters that reduce the
temporal variation of $\q24$ to a minimum and fit the data. 
%
\titlea{Constraints on starburst parameters 
by comparison with the observations}
\titleb{Young starburst (phase I): Fast increase of the magnetic
field?}
The main reason for the high model value of $\q24$ during the early phase
is a lack of synchrotron radiation due to high inverse Compton losses.
The radio emission during the first $\sim 2\times 10^7$ yr is 
therefore -- if the magnetic field remains constant -- 
practically completely thermal (Fig. 6). This
has not been observed, however: Van den Driel et al. 
(1991) determined 
the thermal radio fraction of a sample of starburst galaxies
and derived only $40\%$ at 5 GHz (corresponding to 
about $30\%$ at 2.4 GHz).
A possibly very high thermal fraction should furthermore be noticeable in a
flat radio spectrum ($\prad(\nu)\propto \nu^{-0.1}$) 
which has not been observed either.

The suppression of the synchrotron emission 
will be substantially diminished if the magnetic
field increases
rapidly, already in phase I.  
The behaviour of the magnetic field in a starburst has not been well studied 
yet, but it seems plausible that the field increases.
Ko \& Parker (1989)
proposed a model in which the increase of interstellar turbulence
due to a high SFR activates a turbulent galactic dynamo. 
An alternative process to increase the magnetic field
in a starburst could be the compression by SN shocks of
the ionized gas  in which the field is frozen.
A third field enhancement process could be the systematic dynamo effect
due to a commencing galactic wind as we may be observing in M~82 or 
even at the centre of our Galaxy.
\figeight
\figpos
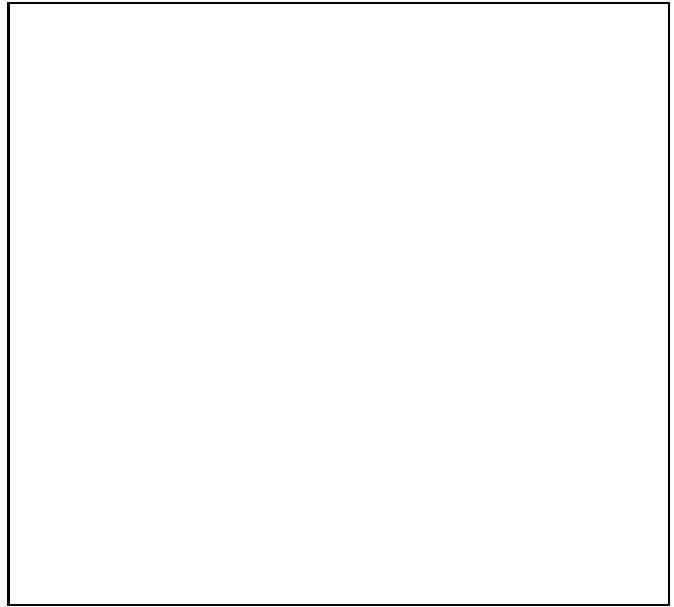
\figtext{8}{$\q24$ is shown for $\tausb=\taugrow=10^7$ yr, $a=100$
(dashed line), $\tausb=\taugrow=3\times 10^7$ yr, $a=50$ (full line)
(both with $\mlow=1 \ms$) and for
$\tausb=\taugrow=3\times 10^7$ yr, $a=50$, $\mlow=3\ms$(dotted line).
An increase of the magnetic field according to 
Eq. (9) has been taken into account in all three cases.}
\endfig
In each of the three scenarios the increase of the magnetic field
is ultimately caused by the star formation.
After the end of the starburst, the magnetic field enhancement
stops and the field dissipates slowly.
Therefore we will tentatively parametrize the temporal variation of
the magnetic field as:
$$B(t)^2\cases{\propto \lfir(t) & {\rm if} ${\partial \lfir\over \partial t}
     > 0$ \cr
     {= \rm constant} & {\rm otherwise}\cr }\eqno\autnum $$ 

In Figs. 8 and 9 the results are shown. It can be seen that the 
magnetic field increase lowers substantially the value of 
$\q24 $ in phase I and allows to fit the data. 
\fignine 
\figpos
\figtext{9}
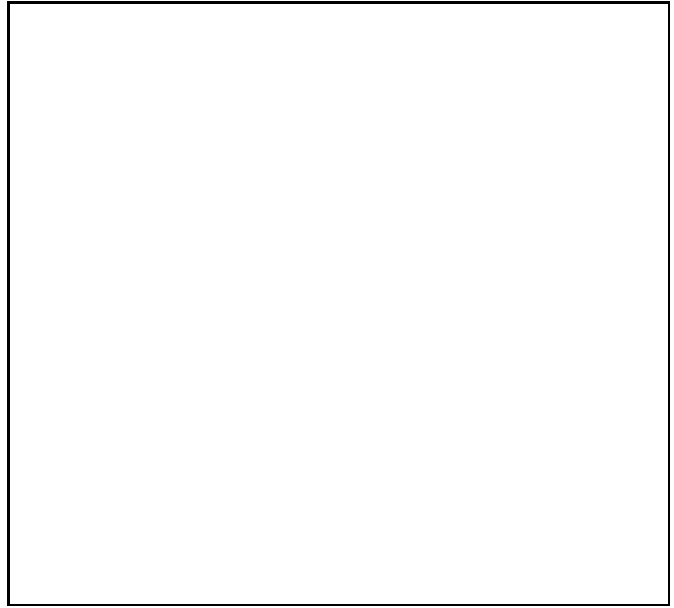
{In the same way as in Fig. 7 the model results and  the
data are compared. The parameters for the model are:
$\taugrow=\tausb=3\times 10^7$ yr, $a=50$, $\mlow=1 \ms$, $A_V=1.5$
(full line). In the dashed line the result for $A_V=5$ 
is shown in order to illustrate the influence
of the extinction on the parameter SB.}
\endfig
We would like to stress that the results presented  do
not crucially depend on the exact parametrization of 
the magnetic field increase during phase I.
The important point is a fast increase of the magnetic field.
The time-scale of this increase  has to be rather
short ($\approx 10^7$) because a large magnetic field strength is
required already in an early stage of the  starburst.
The time-scale for the dynamo activity has been estimated to be
between $10^8$ (Ko \& Parker 1989) and $>10^9$ yr
(Field 1993). Therefore a  dynamo may be a too slow process
for a starburst.
However the alternative processes, such as outward ''combing"
of the field by a mass outflow or of field compression by SN
shocks, would occur on the time-scale of stellar wind bubble
evolution and SN activity. At least the wind activity is 
directly coupled to the FIR enhancement with delays of the
order of a million years only.
An example for this might be the nearby starburst galaxy M~82. It
lies on the FIR/radio correlation (e.g. Wunderlich et al. 1987).
Bernl\"ohr (1992) estimated that the starburst in M~82 has
been active for $1- 5 \times 10^7$ yr, i.e. 
M~82
is in our phase I. The fast galactic outflow from
the starburst nucleus and the corresponding convective electron
transport is consistent with the observed synchrotron emission
and the SN rate (V\"olk et al. 1989).

Parker (1992) has argued that galactic dynamos can be driven by the 
pressure of cosmic rays and the subsequent reconnection of azimuthal field 
lines to produce poloidal fields. This process is competitive to our 
assumed extension of magnetic field lines to ``infinity" by a wind. The 
point here is that also the above version of a galactic dynamo can more or 
less adiabatically follow the time variations in the production of cosmic 
rays, presumably in SNRs, as our data indeed suggest.

\titleb{Evolved starburst (phase II): Time-scale}
If the time-scale of the variation of the star formation rate
is very short, i.e. roughly 
shorter than the time-scale for SN-explosions ($\sim 10^7$ yr, i.e.
the life-time of SN progenitors), the maximum of the attendant
synchrotron emission occurs at a time when
the FIR emission, which has about the same time-scale as the starburst,
has already decreased significantly. This
results in a very low model value of $\q24$ (less than 3 $\sigma$ below
the steady state value).
Therefore, in order to fit the data satisfactorily 
during this phase,  $\taugrow+\tausb$
has to be 
at least comparable to
the SN  time-scale, i.e. about
$10^7$ yr. 
Such time-scales are short enough to allow for repetitive bursts
of star formation on time-scales of 10$^8$ yr as discussed by
Kr\"ugel \& Tutukow (1993).
%
%
Our model results provided 
good fits to the data for 
time-scales longer than about $3 \times 10^7$ yr. 
These numbers are in 
agreement with other authors who found starburst time-scales
between $2\times 10^7$ and some $10^8$ yr (Rieke et al. 1980;
van den Broek et al. 1991; Bernl\"ohr 1993).
We furthermore inferred a minimum time-scale for the increase
of the SFR of $\taugrow\gsim 3\times 10^7$ yr. This slow increase
was necessary 
in order to provide a large enough magnetic field strength and 
SN rate in
the early starburst phase.

\titleb{Post-starburst (phase III): A higher low-mass cut-off of the
IMF?}
We also tested the effect of changes in the IMF, both of the
slope $k$ and the low-mass cut-off $\mlow$.
These two parameters are suggested by many authors to
be different in a starburst in the sense that the
star formation is likely to be biased towards massive stars
(e.g. Rieke et al. 1980; Rieke et al. 
1993; Bernl\"ohr 1992; Wright et al. 1988).
For our model, the most noticeable influence of the IMF occurs in 
the post-starburst phase (III).
When the canonical IMF (Eq. (5)) is assumed for the starburst,
the enhancement in the FIR emission lasts significantly 
longer than the radio emission because of the dust-heating due to
the A stars ($\sim 2 \ms$) produced in the burst.
This is shown by a relatively high model value of $\q24$ in this phase 
(Figs. 4 and 8).
A better agreement with the data can be obtained if
a higher low-mass cut-off, i.e. $\mlow\sim3\ms$ (see Fig. 8), 
is adopted. A flatter slope will have a similar effect. 
Hence, our result seems also in favour of a top-heavy
IMF for starbursts. 

\titlea{Summary and Conclusions}

We have analysed and compared the FIR and radio data of 
a sample of interacting/Markarian galaxies (starburst sample) and a sample
of normal galaxies (control sample) in order to test for the influence
of the star formation activity on the FIR-to-radio ratio.
A model for the FIR and the radio emission in a starburst
was used to interprete the observational 
results and to derive constraints on starburst parameters.
The  following main conclusions could be drawn:
\medskip
\item{--} The mean FIR-to-radio ratio, $\q24$,
showed no sig\-ni\-fi\-cant  difference between the
starburst sample and the control sample 
and is independent of the starburst strength.
\par
\item{--} The magnetic field is required to increase strongly on 
short time-scales ($\approx 10^7$ yrs) in order
to maintain a sufficiently constant $\q24$ in the first 
$\approx 2 \times 10^7$ yr of a starburst.
Otherwise the inverse Compton losses during this phase would
lead to a very low
synchrotron emission, resulting in
a value of $\q24$ significantly higher than the
observed one.
\par
\item{--} The time-scale of the variation of the star-formation rate 
has to be longer than some $10^7$ yr. Otherwise the different time-scales
of the FIR and the radio emission produce large fluctuations  in $\q24$.
\par
\item{--} 
Our results favour a top-heavy
IMF for starbursts (i.e. either a higher lower mass cut-off of
$\mlow\sim 3\ms$ or a flatter slope of the IMF) because
this reduces the contribution of A stars to the dust heating,
thus shortens the time-scale of the FIR emission and 
therefore improves the agreement with the data for
the post-starburst (phase III).
%
\par
\acknow{We would like to thank M. Ko for helpful discussions.
UL gratefully acknowledges the receipt of a stipend of the 
Max-Planck-Gesellschaft and of the Deutsche Forschungsgemeinschaft {\nobreak 
(DFG)}.
HJV and CX acknowledge that part of  their work in this paper has
been done in the framework of the
Sonderforschungsbereich 328 (Entwicklung von Galaxien) of the
DFG.}
\appendix{: Synchrotron emission during a starburst}
Assuming that the time-scale of the radiative energy losses is much
shorter than the time-scale of diffusion, which is certainly a 
valid assumption for a starburst where the inverse
Compton losses of the electrons are very large, we can neglect diffusion
and describe the  
temporal evolution of the electron particle density $f(t,E)$ by the 
following equation:
$${\partial f(t,E)\over \partial t}
        = \biggl({E\over mc^2}\biggr)^{-x} q_{\rm SN}\nu_{\rm SN}(t)+
    {\partial\over \partial E} \biggr\{b(t)
     \, E^2 \, f(t,E)\biggr\} \eqno(A1)$$
where $(E/mc^2)^{-x} q_{\rm SN}\nu_{\rm SN}(t)$ is the source
spectrum of the relativistic
electrons with $\nu_{\rm SN}(t)$ as the supernova rate and $q_{\rm SN}$ as
the number of relativistic electrons produced per supernova and per
energy interval. The power law index of the source spectrum is taken as
$x=2.2$ (V\"olk et al. 1988).
The energy losses $b(t)\propto c_1(\urad(t)+
\ub(t))$ describe the inverse  Compton and synchrotron losses which
are proportional to the energy density of the radiation field $\urad$
and the energy density of the magnetic field $U_{\rm B}$, respectively.
\par
The solution of Eq. (A1) for the starburst population is given by:
$$\eqalign{ &f_{\rm SB}(E,t)=
\biggl({E\over mc^2}\biggr)^{-x}\cdot
  q_{\rm SN} \cr
&\int^t_{\rm max(t_{start},t-\tauloss(E,t))} \diff t'\,\nu_{\rm SN}^{\rm SB}
(t')
 \biggl( 1-E\int^t_{t'} b(t'')\diff t''\biggr)^{x-2}.\cr}\eqno(A2)$$
Here, $\tau_{\rm loss}(E,t)$ is the time-scale of the radiation
losses of the electrons during the
starburst which is defined by the relation:
$$  E\int^t_{t-\tauloss(E,t)} b(t')dt'=1. \eqno(A3)$$
The boundary condition is that at the beginning of the starburst,
at $t=t_{\rm start}$ (which we approximate here as 
$t_{\rm start}=-4\times \taugrow$), the electron density 
$f_{\rm SB}(E,t_{\rm start})=0$. 
For the background population a similiar solution is derived.
The solution for the electron density
can be understood in the following way:
SNe  taking place between $t-\tauloss(E,t)$ and
$t$ contribute to $f(E,t)$. The
contribution of SNe at a time $t'$ gets a weight corresponding to the
radiation losses that are relevant between $t'$ and $t$. 
The higher the radiation losses, the lower is the number
density of electrons in a particular energy range.
\par
In order to calculate $f(E,t)$ from
Eq. (A2), one has to know $\nu_{\rm SN}(t)$ and $b(t)$,
the latter being proportional to $\urad(t)+\ub(t)$.
The supernova rate $\nu_{\rm SN}(t)$ is calculated directly from the SFR
given in Eqs. (3) and (5).
The temporal evolution of the 
energy density of the radiation field, $\urad(t)$, is 
proportional to the bolometric luminosity $L_{\rm bol}(t)$ of the galaxy 
if the volume emitting the radition 
does not change significantly during the starburst:
$$\urad(t)=L_{\rm bol}(t)\cdot
{\urad^{\rm bg}\over L_{\rm bol}^{\rm bg}}\eqno(A4)$$
We take  the radiation energy
density of the steady state galaxy before the starburst 
$\urad^{\rm bg}=1 {\rm eV/cm^3}$, according to the value
in our Galaxy. $L_{\rm {bol}}^{\rm bg}$, its bolometric luminosity,
is derived from the model, assuming
a constant SFR.
\par
The temporal evolution of the magnetic field is 
parametrized according to Eq. (9). 
The energy density
of the magnetic field in a steady state is taken 
as the Galactic value: $U_{B}^{\rm  bg}=1 {\rm eV/ cm^3}$
which corresponds to a magnetic field of $B^{\rm bg}=6.3 \mu$G.
\par
Knowing the electron particle density $f(E,t)$, we can calculate the
synchrotron radiation by:
$$\psyn(\nu,t) \diff\nu=f(E(\nu),t) {\diff E\over \diff t}\bigg|_{\rm syn}
{\diff E\over \diff\nu} \diff \nu \eqno(A5)$$
with $(\diff E / \diff t)|_{\rm syn}$
as the energy loss by synchrotron radiation and  $\diff E /\diff \nu$
given through:
$$\nu=\biggl({E\over m c^2}\biggr)^2 {\nu_{\rm G}(B)\over 0.45}\eqno(A6)$$
(Longair 1994) with $\nu_{\rm G}$ being the gyrofrequency.
This corresponds to 
the simplification that an
electron of energy $E$ emits the whole synchrotron radiation at the
maximum frequency of the synchrotron spectrum.
\par
In order to find the absolute normalization of the synchrotron radiation
we scale the synchrotron radiation such that
the FIR/radio ratio $\q24$ for the background population is equal to the
observed average value of the control sample.
\par
\begref{References}
\ref
Aaronson M., Huchra J., Mould J., Schechter P., Tully R.B., 1982,
 \apj 258, 64
\ref
Balzano V.A., 1983, ApJ 268, 602
\ref Basu S., Rana R.C., 1992, \apj 393, 373 \par
\ref Bernl\"ohr K., 1992, \aa 263, 54 \par
\ref Bernl\"ohr K., 1993, \aa 268, 25 \par
\ref Binggeli, B. Sandage A., Tammann G. A., 1985, \aj 90, 1681
\ref
Bothun G.D., Lonsdale C.J., Rice, W., 1989, ApJ 341, 129
\ref Condon J.J., 1992, \araa  30, 575 \par
\ref Condon J.J., Huang Z.P. Yin Q.F. 1991, \apj 378, 65\par
\ref Cox M.J., Eales S.A.E., Alexander P., Fitt A.J., 1988, \mnras
    237, 381 \par
\ref Cox P., Kr\"ugel E., Mezger P.G., 1986, \aa 155, 155 \par
\ref
de Vaucouleurs G., de Vaucouleurs A., Corwin H. et al., 1991,
 Third Reference Catalog of Galaxies. New York, Springer-Verlag (RC3)
\ref
Dressel L.L., Condon J.J. 1978, ApJS 36, 53
\ref
Feigelson E.D., Nelson P.I. 1985, ApJ 293, 192
\ref Field G., 1993, in: J. Franco, F. Ferrini, G. Tenorio-Tagle (eds.), 
Star Formation, Galaxies and the Interstellar Medium.
Cambridge University Press, Cambridge 
\ref
Franceschini A., Danese L., De Zotti G.,  Toffolatti L., 1988a,
 MNRAS 233,157
\ref
Franceschini A., Danese L., De Zotti G.,  Xu C., 1988b,
 MNRAS 233,157
\ref
Gavazzi G., Boselli A.,  Kennicutt R., 1991, AJ 101, 1207
\ref G\"usten R., Mezger P.G., 1983, Vistas in Astronomy 26, 159
\ref Helou G., Soifer B.T., Rowan-Robinson M., 1985, ApJ 298, L7
\ref
Helou G. Khan I.R., Malek L.,  Boehmer L., 1988, ApJS 68, 151
\ref Huchra J.P., 1977, \apj 217, 928 \par
\ref
Kennicutt R.C., 1983, \apj 272, 54
\ref
Kennicutt R., Keel W., van der Hulst J., Hummel E., Roettiger K., 1987,
 AJ 93, 1011
\ref Ko C.M., Parker E.N., 1989, \apj 341, 828 \par
\ref Kr\"ugel E., Tutukov A.V., 1993, \aa 275, 416 \par
\ref
Larson R.B., Tinsley B.M., 1978, ApJ 219, 46
\ref Lisenfeld U., V\"olk H.J., Xu C., 1996, \aa 306, 677\par
\ref Longair M.S., 1994, High Energy Astrophysics. Vol 2.,
Cambridge University Press
\ref
Mazzarella J.M., Balzano V.A., 1986, ApJS 62, 751
\ref
Mazzarella J.M., Bothen G.D., Borrosen T.D., 1991, AJ 101, 2034
\ref
Nilson P., 1973, Uppsala General Catalogue of Galaxies. Uppsala Astronomical
 Observatory, Uppsala (UGC)
\ref Parker E.N., 1992, \apj 401, 137 \par
\ref Rieke G.H., Lebofsky M.J., Thompson R.I., Low F.J., Tokunaga T.,
1980, \apj  238, 24 \par
\ref Rieke G.H., Loken K., Rieke M.J., Tamblyn P., 1993, \apj 412,
99 \par
\ref Scalo J.M., 1986, Fundam. of Cosmic Research,  11 \par
\ref
Schmitt J.H.M.M., 1985, ApJ 293, 178
\ref
Telesco C., Wolstencroft R., Done C., 1988, ApJ 329, 174
\ref
Thuan T.X., Sauvage M., 1992, A\&AS 92, 749
\par
\ref van den Broek A.C., de Jong T., Brink K. 1991 \aa  246, 313 \par
\ref van den Driel W., van den Broek A.C., de Jong T., 1991
   \aas 90, 55 \par
\ref
V\"olk H.J., Xu C. 1994, Infrared Phys. Technol. 35, 527 \par
\ref V\"olk H.J., Zank L.A., Zank G.P, 1988, \aa 198, 274 \par
\ref V\"olk H.J., Klein U., Wielebinski R., 1989, \aa 213, L12 \par
\ref Wright G.S., Joseph R.D., Robertson N.A., James P.A., Meikle
W.P.S., 1988, \mnras  233,1 \par
\ref Wunderlich E., Klein U., Wielebinski R., 1987, \aas  69, 487 \par
\ref 
Xu, C., 1990, \apj 365, L47 \par
\ref Xu C., Buat V., 1995, \aa 293, L65 \par
\ref
Xu C.,  De Zotti G. 1989, A\&A 225, 12
\ref
Xu C.,  Sulentic J.W. 1991, ApJ 374, 407
\ref Xu C., Lisenfeld U., V\"olk H.J., Wunderlich E., 1994, 
\aa 282, 19 \par

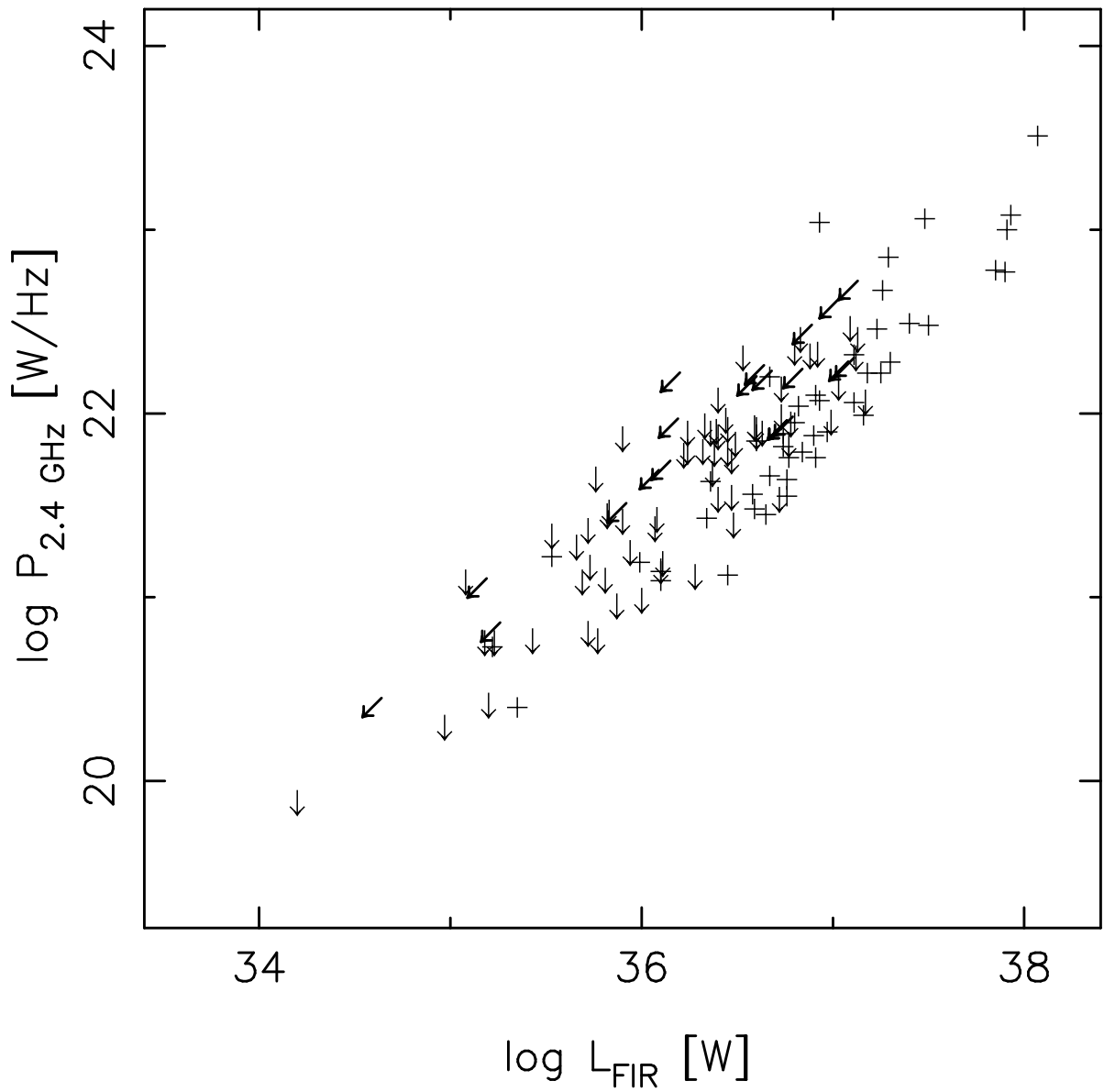
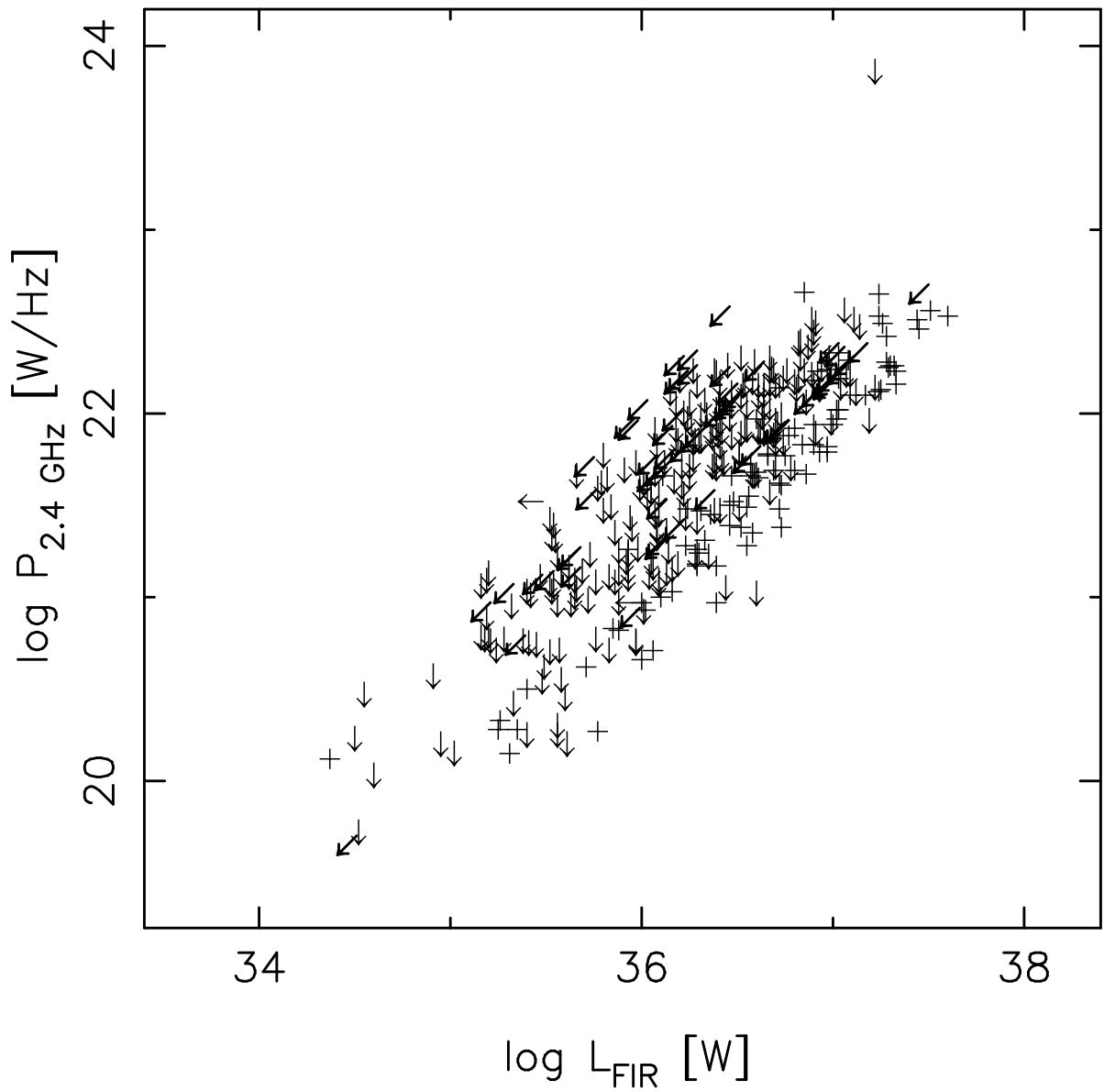
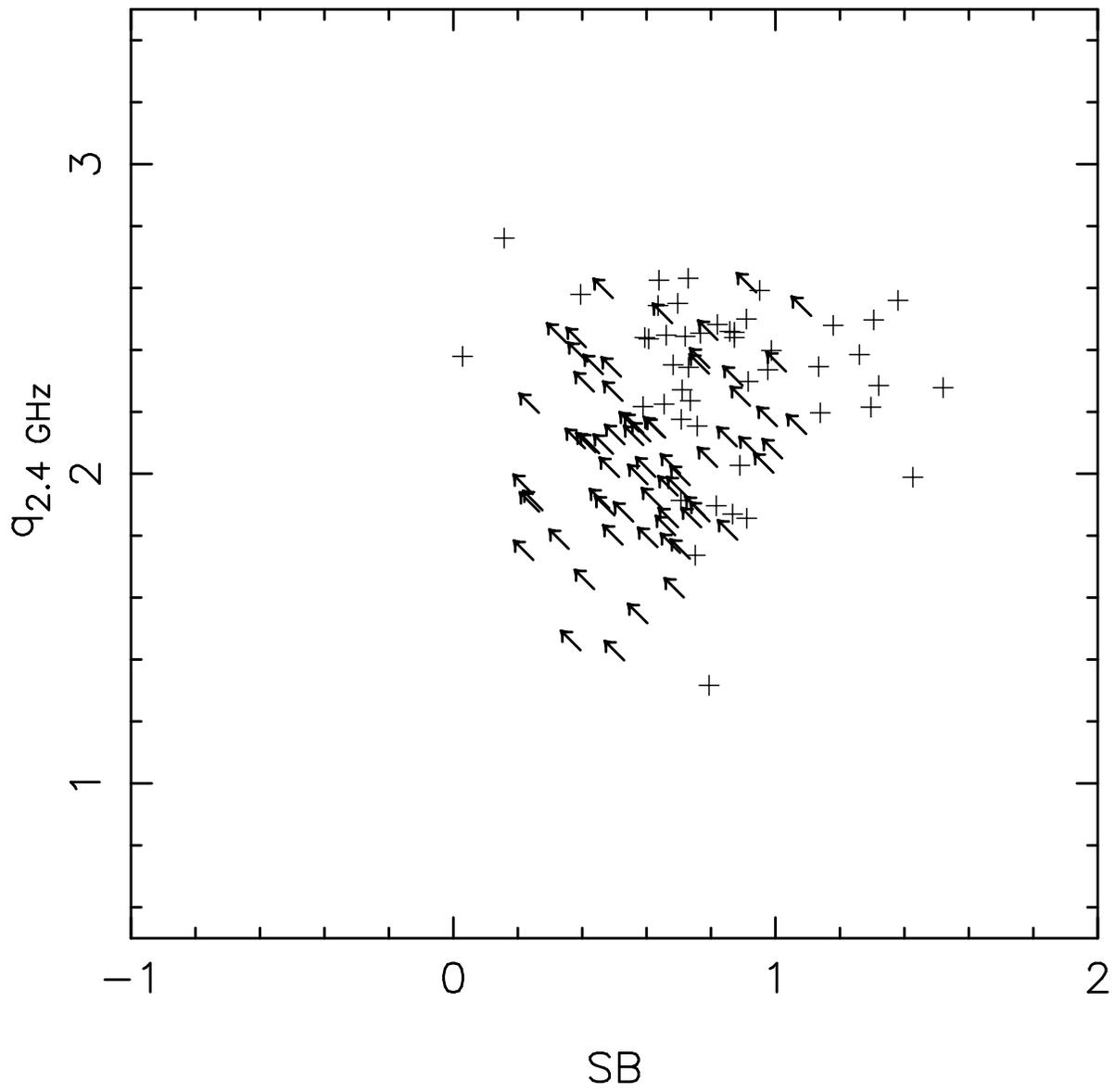
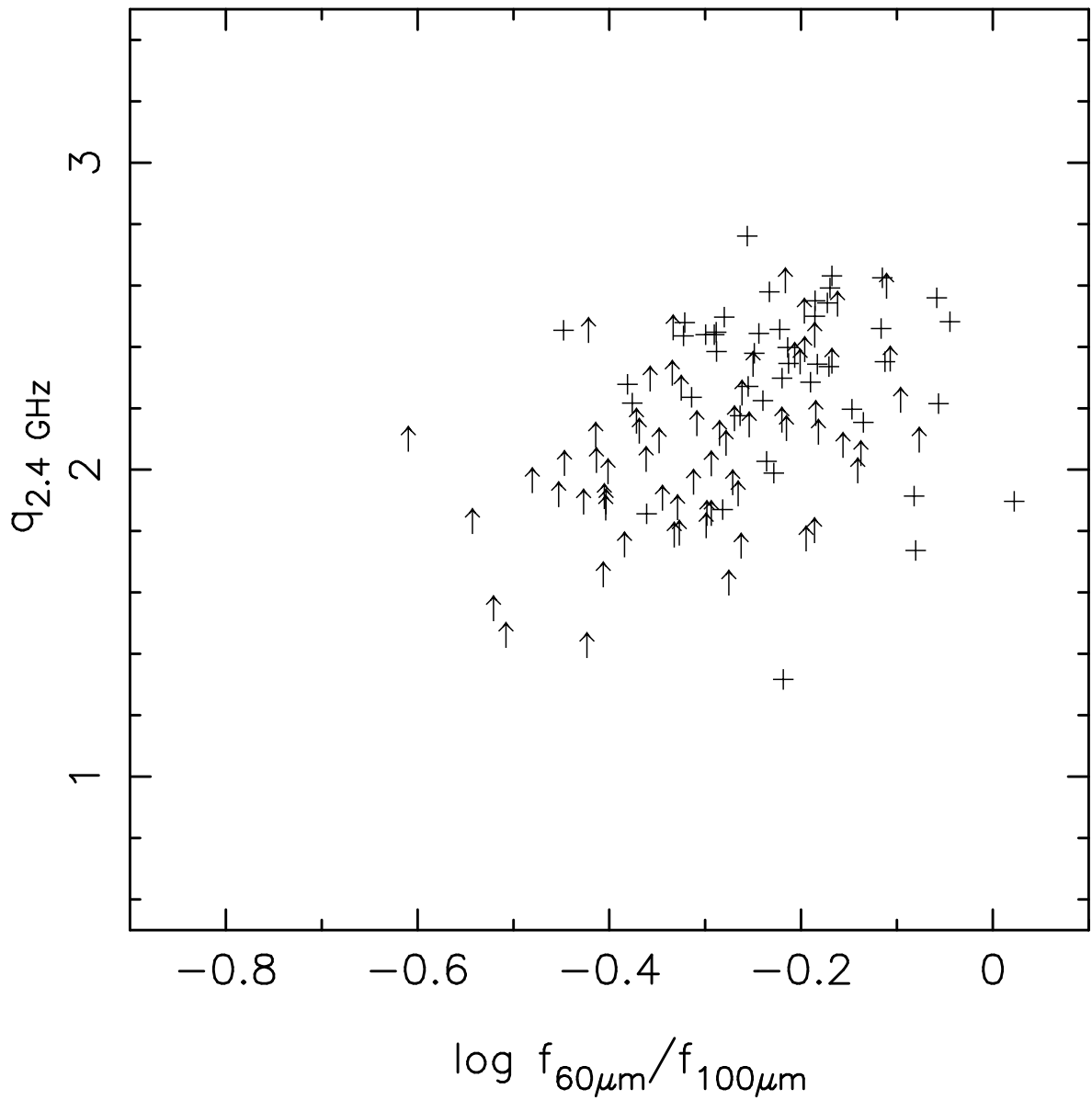
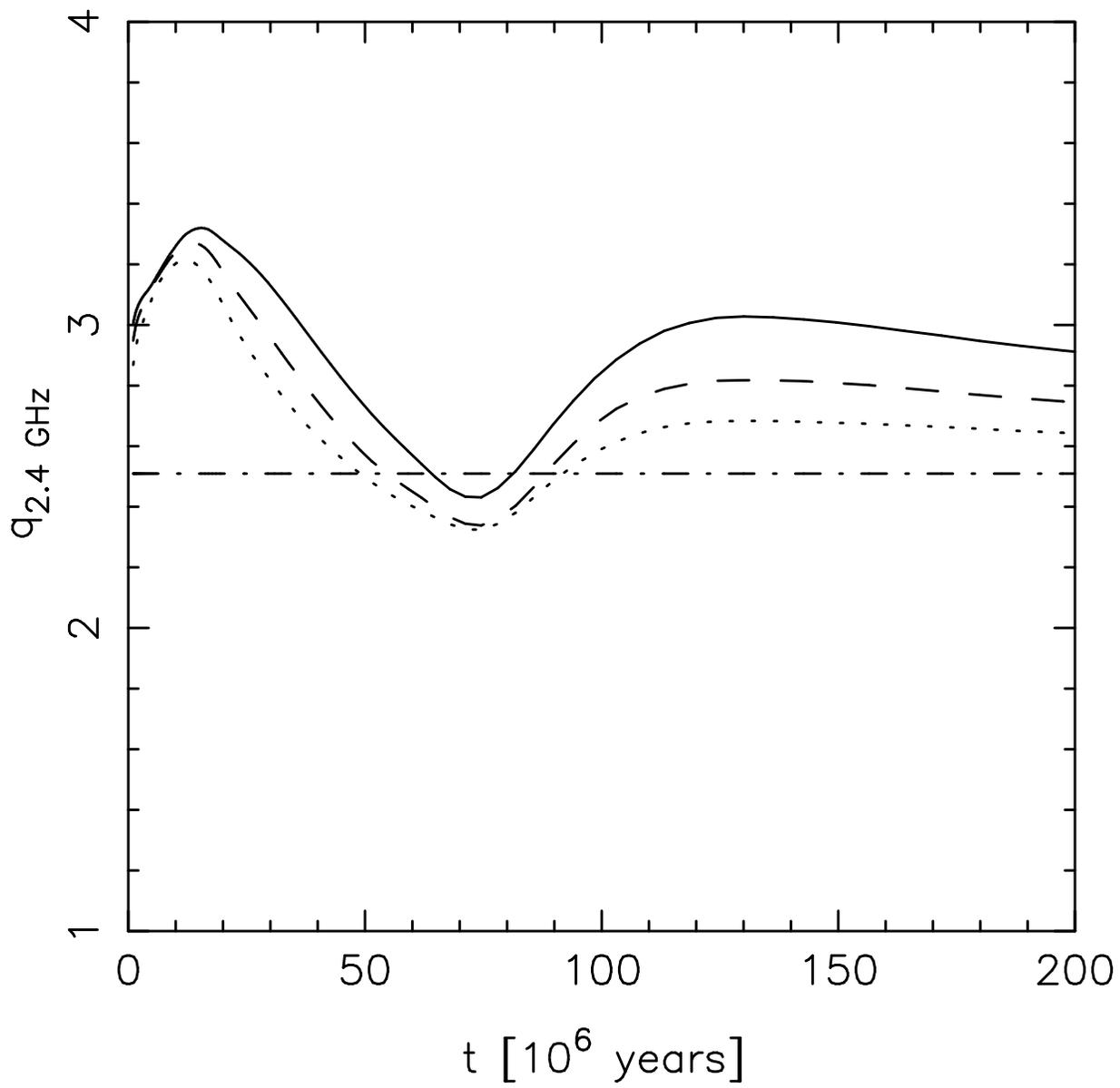
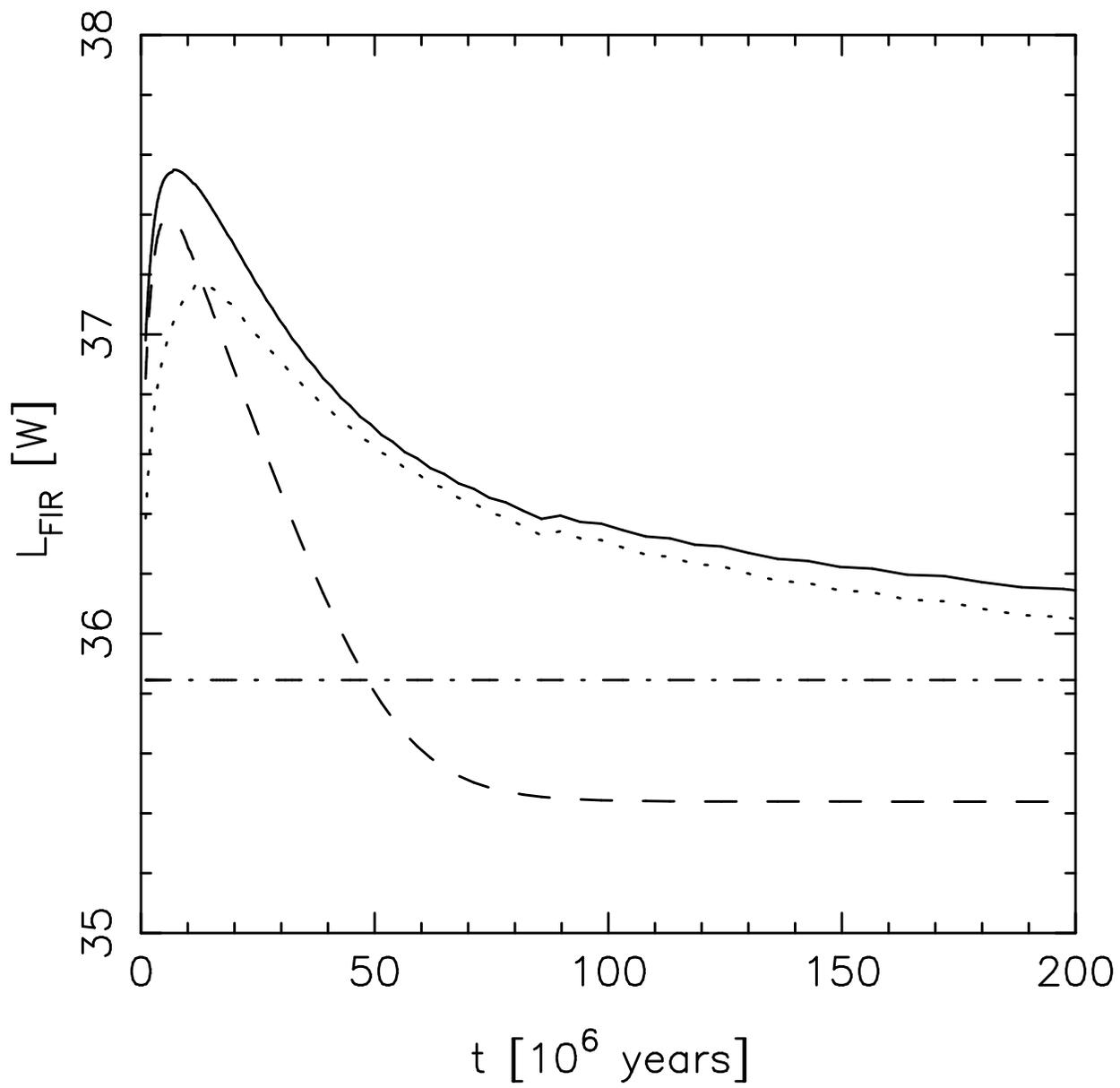
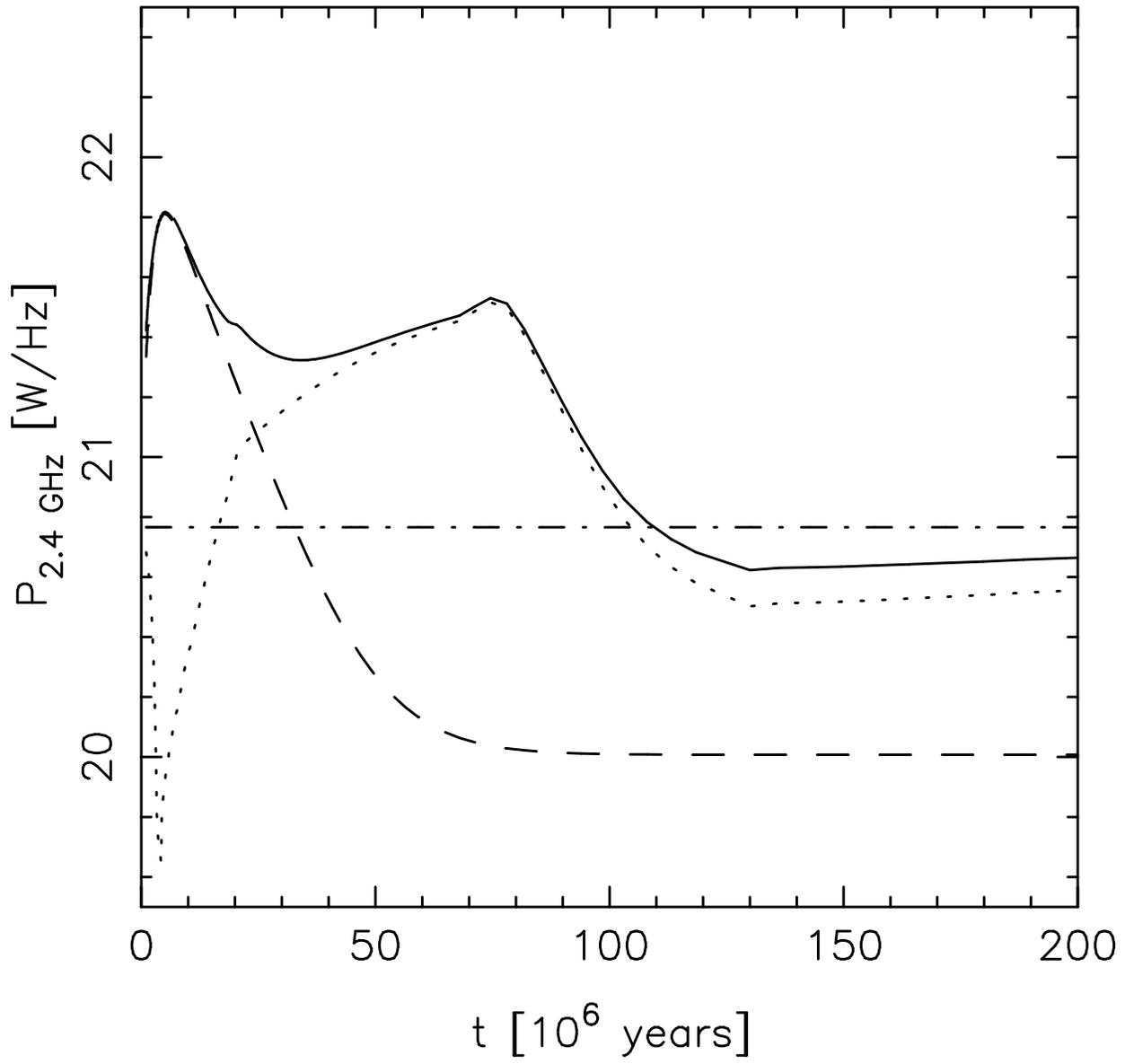
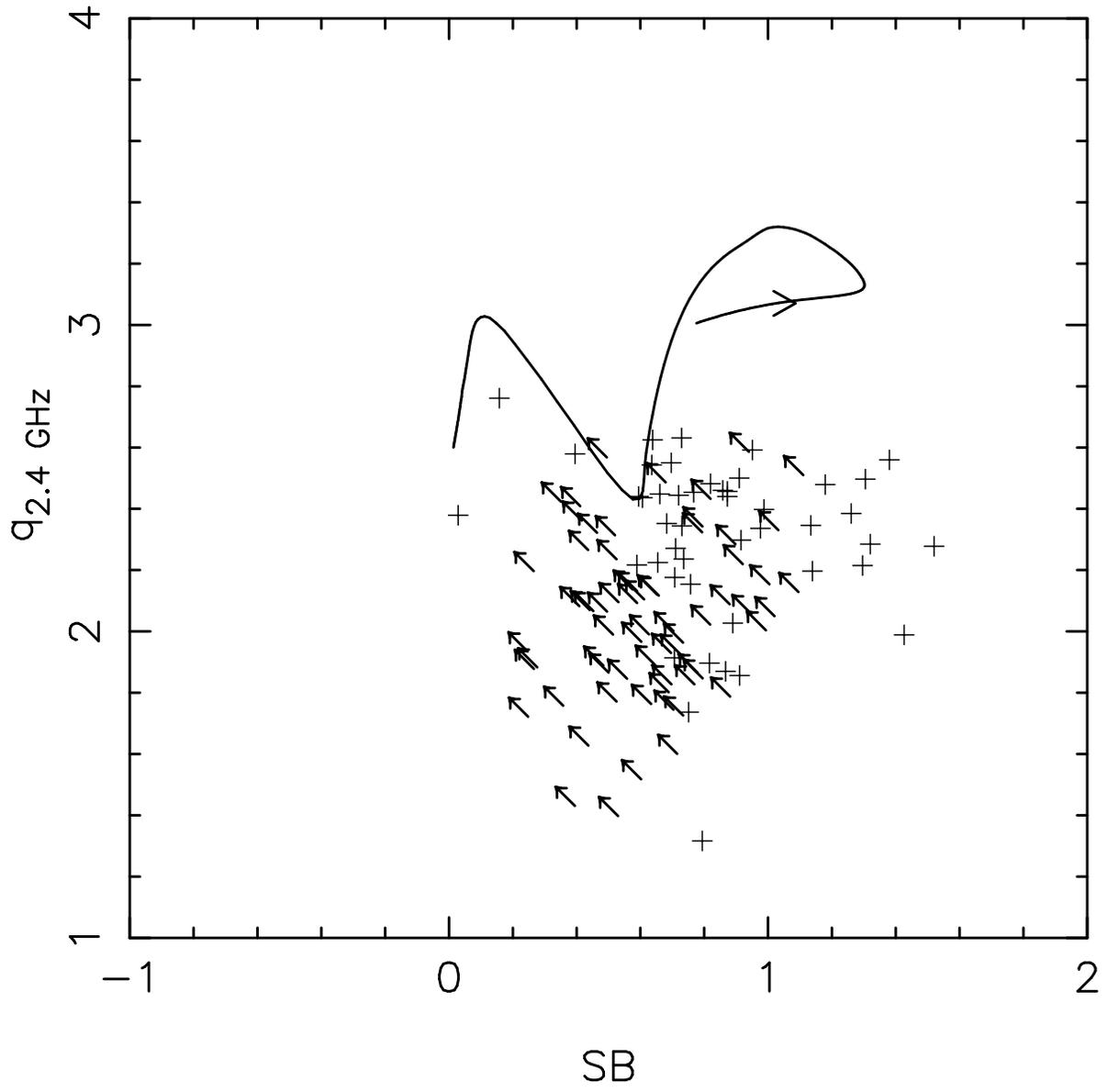
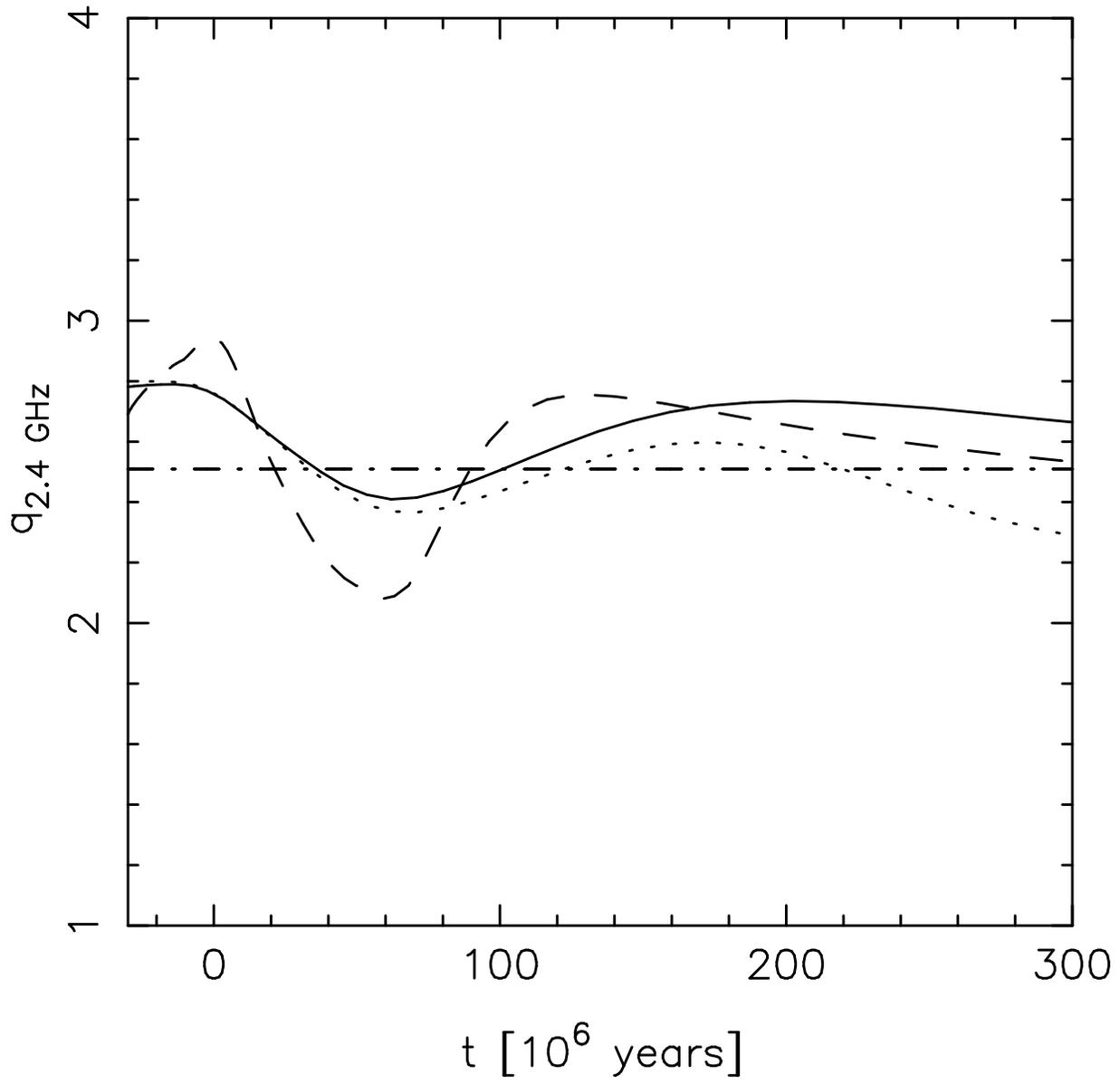
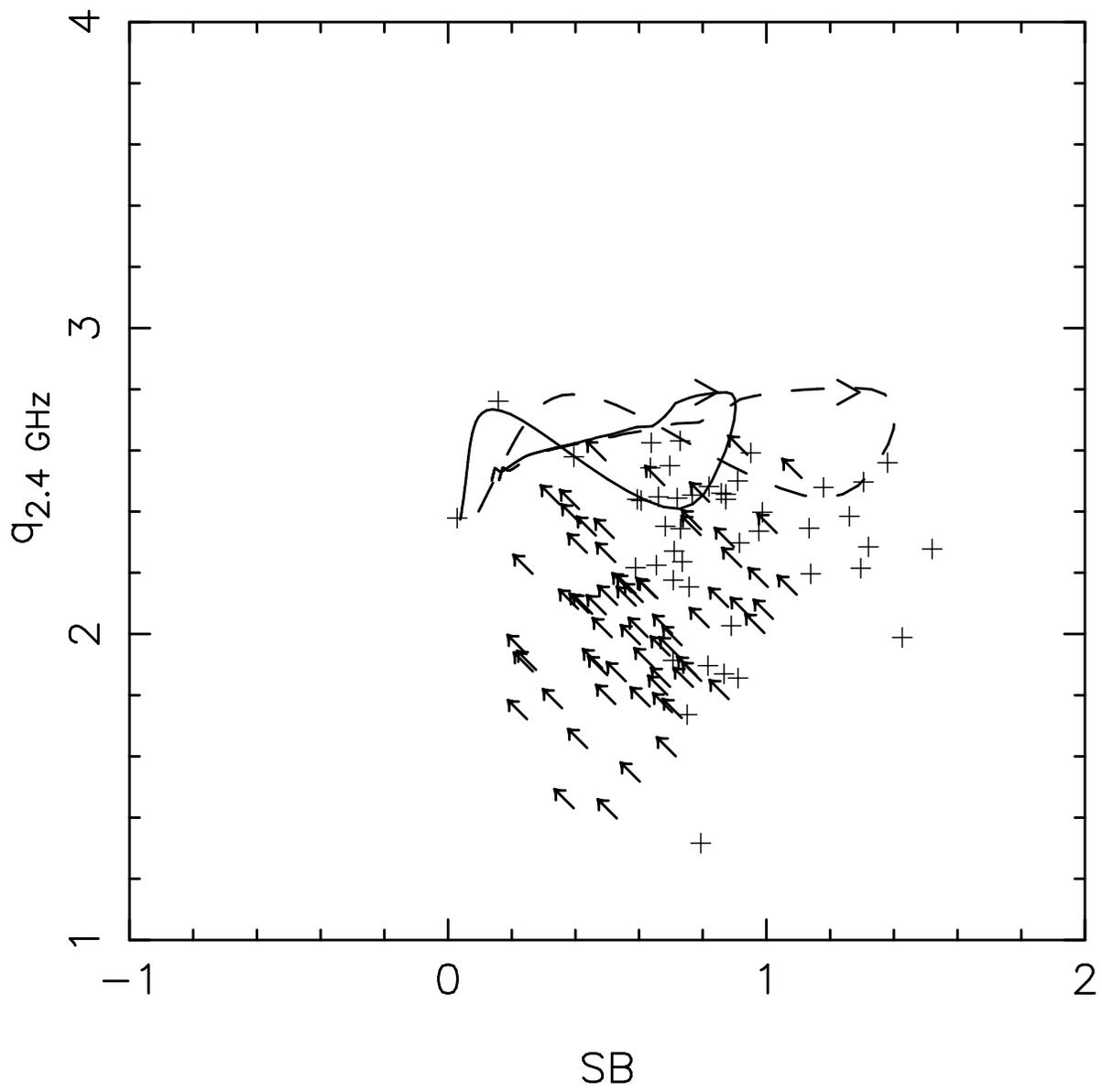
\end